\documentclass[referee]{aa} % for a referee version
\usepackage{pdflscape}
\usepackage{longtable}
\usepackage{rotating}
\usepackage{booktabs}
\usepackage{amsmath}
\usepackage{gensymb}

\usepackage{xcolor}
\usepackage{graphicx}
\usepackage{url}
\usepackage{amsmath}
\usepackage[colorlinks=true,linkcolor=magenta,citecolor=blue,filecolor=cyan,urlcolor=purple,final=true]{hyperref}
\newcommand{\jswsc}{   {Journal of Space Weather and Space Climate}}
\newcommand{\natco}{   {Nature Communications}}

\usepackage{txfonts}
\begin{document}

   \title{Effects of optimisation parameters on data-driven magnetofrictional modelling of active regions}

   \author{A. Kumari \inst{1,2}
        \and
        D. J. Price \inst{1} 
         \and
         F. Daei \inst{1}
         \and
         J. Pomoell \inst{1}
        \and
         E. K. J. Kilpua \inst{1}
          }

   \institute{Department of Physics, University of Helsinki, P.O. Box 64, FI-00014, Helsinki, Finland \\
              \and
   NASA Goddard Space Flight Center, Greenbelt, MD 20771, USA \\
   \email{anshu.kumari@helsinki.fi}\\
             }

  % \date{Received September 15, 1996; accepted March 16, 1997}

% \abstract{}{}{}{}{}
% 5 {} token are mandatory
 
  \abstract
   {The solar magnetic field plays an essential role in the formation, evolution, and dynamics of large-scale eruptive structures in the corona. Estimation of the coronal magnetic field, the ultimate driver of space weather, particularly in the ‘low’ and ‘middle’ corona, is presently limited due to practical difficulties. Data-driven time-dependent magnetofrictional modelling (TMFM) of active region magnetic fields has been proven to be a useful tool to study the corona. The input to the model is the photospheric electric field that is inverted from a time-series of the photospheric magnetic field. Constraining the complete electric field, i.e., including the non-inductive component, is critical for capturing the eruption dynamics. We present a detailed study of the effects of optimisation of the non-inductive electric field on TMFM of AR12473.}
   { We study the effects of varying the non-inductive electric field on the data-driven coronal simulations, for two alternative parametrisations.
   By varying parameters controlling the strength of the non-inductive electric field, we explore the changes in flux rope formation and their early evolution and other parameters, e.g., axial flux and magnetic field magnitude.
   }
   { We use the high temporal and spatial resolution cadence vector magnetograms from the Helioseismic and Magnetic Imager (HMI) onboard the Solar Dynamics Observatory (SDO).
   The non-inductive electric field component in the photosphere is critical for energizing and introducing twist to the coronal magnetic field, thereby allowing unstable configurations to be formed. We estimate this component using an approach based on optimising the injection of magnetic energy.  
}
   {Our data show that flux ropes are formed in all of the simulations except for those with the lower values of these optimised parameters.
   However, the flux rope formation, evolution and eruption time varies depending upon the values of the optimisation parameters. The flux  rope  is  formed and  has  overall  similar  evolution  and  properties  with  a  large range of non-inductive electric fields needed to determine the non-inductive electric field component that is critical for energizing and introducing twist to the coronal magnetic field.
   %Did you get FR? E= 0 no FR, omega/U FR 
   }
   {This  study  shows  that irrespective  of non-inductive electric field values,  flux  ropes  are  formed  and  erupted, which indicates that data-driven TMFM can be used to estimate flux rope properties early in their evolution without needing to employ a lengthy optimisation process.}


\titlerunning{Effects of optimisation parameters on MFM of ARs}
\authorrunning{Kumari et al.}
   \maketitle
%
%-------------------------------------------------------------------

\section{Introduction}
\label{sec:section1}

 Coronal Mass Ejections \citep[CMEs; e.g.,][]{Webb2000} are of primary interest for space weather as they cause the largest disturbances in the near-Earth space environment \citep[e.g.,][]{Kilpua2017a,Zhang2007}. To forecast their impacts on our planet requires modelling their properties and evolution from the Sun to the Earth. Both semi-empirical and physics-based space weather models require realistic input parameters to constrain their magnetized CME models \citep[e.g.,][]{Kilpua2019}. In particular, recent advancement includes runs with spheromaks and flux rope CMEs with 3-D magnetohydrodynamic heliospheric models such as the EUropean Heliospheric FORecasting Information Asset \citep[EUHFORIA; e.g.,][]{Pomoell2018,Scolini2019,Scolini2020,Verbeke2019,Asvestari2021} and Space-weather-forecast-Usable System Anchored by Numerical Operations and Observations  \citep[SUSANOO; e.g.][]{Shiota2016}
 %and ENLIL \citep{Odstrcil1999,Odstrcil2002}.
 The goal is to predict a time series of magnetic field vectors impinging the Earth with good accuracy.
% Work in progress
% intrinsic CME flux rope params
%

Currently, the magnetic field in the corona is difficult to routinely measure reliably and with global coverage due to the relatively weak field strength and the plasma being hot and tenuous, resulting in the broadening of spectral lines \citep{Lin2000}. The information of intrinsic CME flux rope parameters must thus be obtained by other means, using either indirect observational proxies or data-driven coronal modelling \citep[e.g.,][and references therein]{Kilpua2019, Palmerio2017}. The knowledge of the magnetic field of a CME is not only important for space weather but also for fundamental investigations of the formation, destabilization, and eruption of solar magnetic flux ropes \citep[e.g.,][]{Green2018,Welsch2018,Kilpua2019}.

From a space weather forecasting point of view a data-driven model for eruptive coronal fields has to be computationally fast. Time-dependent magnetohydrodynamic (MHD) simulations are in principle well suited for investigating flux rope dynamics in the corona \citep[e.g.,][]{Jiang2016}, but they are computationally expensive and not all necessary boundary conditions are routinely available from observations. An alternative approach is to simplify or even neglect the thermodynamics, which is justified by the dominance of magnetic forces in the low corona. One such approach, the nonlinear force free field (NLFFF) method, uses photospheric magnetograms to extrapolate a static snapshot of the coronal magnetic field at a particular instance in time \citep{Wiegelmann2012}.

The magnetofrictional method (MFM) \citep{Yang1986} is a particular method for computing NLFFF configurations. The basic principle of the method is the addition of a frictional term $-\nu \mathbf{v}$, where $\nu$ is the frictional coefficient and $\mathbf{v}$ the plasma velocity, to the MHD momentum equation and assuming a low-beta (pressure gradient ignored) and quasi-static evolution. 
%This means that the frictional velocity is added to the MHD momentum equation and the method assumes quasi-static and low-beta situation. 
With these assumptions, the MHD momentum equation reduces such that  an explicit relation for the magnetofrictional velocity is obtained and given as $\mathbf{v} = \frac{1}{\nu} \frac{\mu_0 \mathbf{J} \times \mathbf{B}}{B^2}$ that is further used to evolve the magnetic field according to Faraday's law via an electric field obtained by invoking Ohm's law. The velocity, being proportional to the Lorentz force, evolves the system towards a force-free minimum-energy state in the absence of a net Poynting flux.

MFM can be extended to a time-dependent version using time-varying photospheric boundary conditions \citep{Ballegooijen2000}.
Several studies have demonstrated that data-driven time-dependent magnetofrictional modelling (TMFM) can successfully describe the formation and initial rise of solar flux ropes when using a time-sequence of boundary conditions derived from photospheric observations \citep[][]{Cheung2012,Gibb2014, Fisher2015,Yardley2018,Pomoell2019, Price2019,Price2020,Kilpua2021, Lumme2022}.

The sole driving boundary condition required by TMFM is the photospheric electric field. The so-called inductive electric field component ($\mathbf{E}_I$) is obtained directly by inverting Faraday's law using as input a time series of photospheric magnetograms, while for specifying the remaining non-inductive (curl-free) component %(given as gradient of the scalar potential $\nabla \psi$) 
one has to incorporate additional information, e.g. using Dopplergrams and optical flow methods \citep{Kazachenko2015} or, alternatively, \textit{ad hoc} prescriptions \citep{Lumme2017, Welsch2018,Yeates2022}. 
%\textbf{Recently, \cite{Yeates2022} described a technique where the authors constrained the non-inductive electric field using sparse matrix.}
%The non-inductive component has been shown to be crucial for accurately capturing the injection of energy and helicity into the corona and as a consequence heavily influences flux rope early dynamics in the TMFM simulation.
%
%Three \textit{ad hoc} assumptions are generally in use. One is setting the inductive electric field simply to zero. 
%
%Others are related to setting it to depend on the spatially homogeneous twisting motion at the photosphere or to the emerging twisting flux to the photosphere \citep{Cheung2012, Cheung2015}. 
%
%In another, the horizonal component of fluid vorticity is a free parameter \citep[$\Omega$ - assumption;][]{Cheung2012, Cheung2015} and in the final case, the vertical speed at which the twisted flux tube rises to the photosphere ($U$-assumption). %
Previous studies have shown that the non-inductive electric field component has a significant contribution to the total electric field and ignoring it estimates only some percent of the Poynting flux \citep[][]{Fisher2010,Kazachenko2014,Price2019}.
The non-inductive component has also been found to be critical for the formation and rise of flux rope configurations in magnetofrictional simulations \citep[e.g.,][]{Pomoell2019, Cheung2012}. 

The optimisation method described in \cite{Lumme2017} estimates the non-inductive electric field by employing
\textit{ad hoc} assumptions on the functional form of the sources responsible for generating the electric field. In the process, free parameters in the non-inductive electric field are introduced, and are constrained 
%the values of free-parameters, i.e. $U$ and $\Omega$ 
by finding the best match with the energy injection computed by  \citep[Differential Affine Velocity Estimator for Vector Magnetograms;][]{Schuck2008}. While well defined, this approach 
%however not the "ground truth" for determining free parameters $U$ and $\Omega$. 
yields an electric field based on a single metric while ignoring other possible metrics, such as the injection of helicity or the behaviour of the coronal magnetic field when used to drive a coronal model.
%For example, \cite{Kilpua2021} obtained a flux rope and its coherent rise using twice the energy-optimised value. 

It is an open question how sensitive the formation and evolution of coronal flux ropes and their key parameters are to the 
%values of $U$ and $\Omega$. 
choices made in estimating the photospheric non-inductive electric field. 
%For space weather forecasting purposes, a key focus is to get a realistic estimate of flux rope parameters. 
%I.e., if the flux rope properties do not change significantly with the value of free parameters, one could quickly drive TMFM with a preset values of $U$ and $\Omega$ to obtain space weather relevant information.
%It is therefore important to explore how flux rope properties vary depending on the choice of these free parameters.
%
The purpose of this study is to explore how the choice of 
the non-inductive electric field
%$U$ and $\Omega$ 
in the TMFM simulation affect the flux rope formation and early evolution, as well as their key properties. We perform a detailed study of the effects of the driving electric field on AR 12473 during December 22, 2015 to January 02, 2016. \cite{Price2020} performed a detailed study of the magnetic evolution of AR 12473 using TMFM with the energy-optimised non-inductive driving electric field. In the present study, we show how varying the optimisation parameters affects the evolution and eruption of the flux rope. The paper is arranged as follows: In section \ref{sec:section2}, we describe the active region (AR) we have studied and the preparation of the data for TMFM. We give a detailed description of the method used to perform the TMFM. We describe the analysis of the simulated data in section \ref{sec:section3} and provide results obtained. Finally, we
conclude this study in section \ref{sec:section4} with the summary. %and important findings.

\section{Data and Methods}
\label{sec:section2}

\subsection{Event overview}
\label{sec:section2.1}

We studied the evolution of AR 12473 from December 22, 2015 to January 02, 2016. The location of this AR on the aforementioned days was at S22E66\footnote{\url{https://www.solarmonitor.org/index.php?date=20151222&region=12473}} and S21W86\footnote{\url{https://www.solarmonitor.org/index.php?date=20160102&region=12473}} (Heliographic Coordinates) as seen with the Helioseismic and Magnetic Imager \citep[HMI;][]{Scherrer2012} onboard the Solar Dynamics Observatory \citep[SDO;][]{Pesnell2012}. The sunspot associated with the AR was of a bipolar ($\beta$) nature. There were several C and M class flares associated with this AR \cite[eg.][]{Mulay2021}. 
%Ground based radio spectrographs such as e-Callisto and RSTN had also reported many type IV radio bursts associated with this particular AR.

On December 28, 2015 at approximately 11:30 UT an eruption was visible in the multi-wavelength observations from the Atmospheric Imaging Assembly \citep[AIA;][]{Lemen2012} onboard the SDO. On the day of this eruption, the AR was located close to the disk centre (S23W11). An M1.9 class flare was associated with this eruption. The start, peak and end times of this flare were 11:20 UT, 12:45 UT and 15:00 UT, respectively. 
There was a halo CME associated with the eruption as viewed from Earth\footnote{\url{https://cdaw.gsfc.nasa.gov/CME_list/halo/halo.html}}, which was first seen at 01:12 UT by the C2 coronagraph of the Large Angle Spectroscopic Coronagraph \citep{Brueckner1995} onboard on the Solar and Heliospheric Observatory \citep{Domingo1995}. The linear and second-order speed of the CME was $\approx$ 1212 km/s and $\approx$ 1471 km/s, respectively\footnote{\url{https://cdaw.gsfc.nasa.gov/CME_list/UNIVERSAL/2015_12/yht/20151228.121205.w360h.v1212.p163g.yht}}. 
%The coronagraph data indicate an acceleration of $\approx$ 4.6 km/s$^2$.
An interplanetary CME (ICME) was recorded on December 31, 2015 with the shock and leading edge at $\approx$ 00:50 UT and $\approx$ 17:00 UT\footnote{\url{http://www.srl.caltech.edu/ACE/ASC/DATA/level3/icmetable2.htm}}, respectively with the Advanced Composition Explorer \citep[ACE;][]{Chiu1998}.

\subsection{Data}
\label{sec:section2.2}

For the data-driven coronal simulation \citep{Pomoell2019}, we use a time-sequence of disambiguated vector magnetograms obtained by HMI. The vector magnetograms have a spatial and temporal resolution of $\approx 0.5^"$ in the plane of sky and $\approx 700$ s, respectively. 
%These dataset are developed from full-disk filtergrams which are recorded at a 135 s time cadence.
The HMI data were downloaded from the Joint Science Operations Center
(JSOC)\footnote{\url{http://jsoc.stanford.edu/HMI/Vector_products.html}} using the ELECTRIC field Inversion Toolkit (ELECTRICIT) developed by \cite{Lumme2017}. 
%We used the SunPy package in python for downloading the hmi.B\textunderscore720s titled dataset. 

\subsection{Methods}
\label{sec:section2.3}

\subsubsection{Vector magnetogram processing}
\label{sec:section2.3.1}

The data were processed using ELECTRICIT to make them suitable as input for the simulation. For this, firstly, the very strong and weak magnetic field vector pixels ($B \geq 750 ~\textrm{Mx} ~\textrm{cm}^{-2}$ and $B \leq 250 ~\textrm{Mx} ~\textrm{cm}^{-2}$, respectively) were disambiguated using the minimum energy method \citep{Metcalf1994} and the potential field acute angle method \citep{Liu2017}, respectively. The spurious pixels in the vector magnetograms were then removed by checking the 
formal error of total magnetic field strength by the Stokes inversion module \citep{Hoeksema2014}, which was fixed as $\sigma_B = 750 ~\textrm{Mx} ~\textrm{cm}^{-2}$. These spurious pixels were then replaced with the median of their surrounding good pixel values. 

This disambiguated dataset was used to produce a time series of re-projected magnetogram cutouts of the AR from the start to the end date of the observation. Using ELECTRICIT, we tracked AR12473 across the disk and produced time series of cutouts in a coordinate system with the AR remaining in the centre of the cutout.
The tracking is similar to the Space-weather HMI Active Region Patches (SHARPs) \citep{Bobra2014, Hoeksema2014}, which tracks each HMI Active Region Patch (HARP) using a fixed rotation rate. We also masked the noise-dominated weak-field pixels. The data was re-projected using Mercator map projection centered at the region of interest to a local Cartesian system.
%This processed dataset served as the lower boundary condition for the simulation. 

The re-projected time series cutouts were produced from 18:00 UT December 22, 2015 to 22:00 UT January 02, 2016, which included the heliographic longitude from $\approx -50^{\circ}$ to $\approx +50^{\circ}$. The previous and later data were discarded due to the poor quality of magnetograms closer to the limb \citep{Sun2017}. The dataset contained the emergence of AR12473, the evolution of the AR and the eruption which took place on December 28, 2015 at $\approx$ 11:30 UT.
This time series was used for optimising the free parameters of the electric field inversion.

\begin{figure*}[ht]
    \centering
        \includegraphics[width=1\textwidth,clip=]{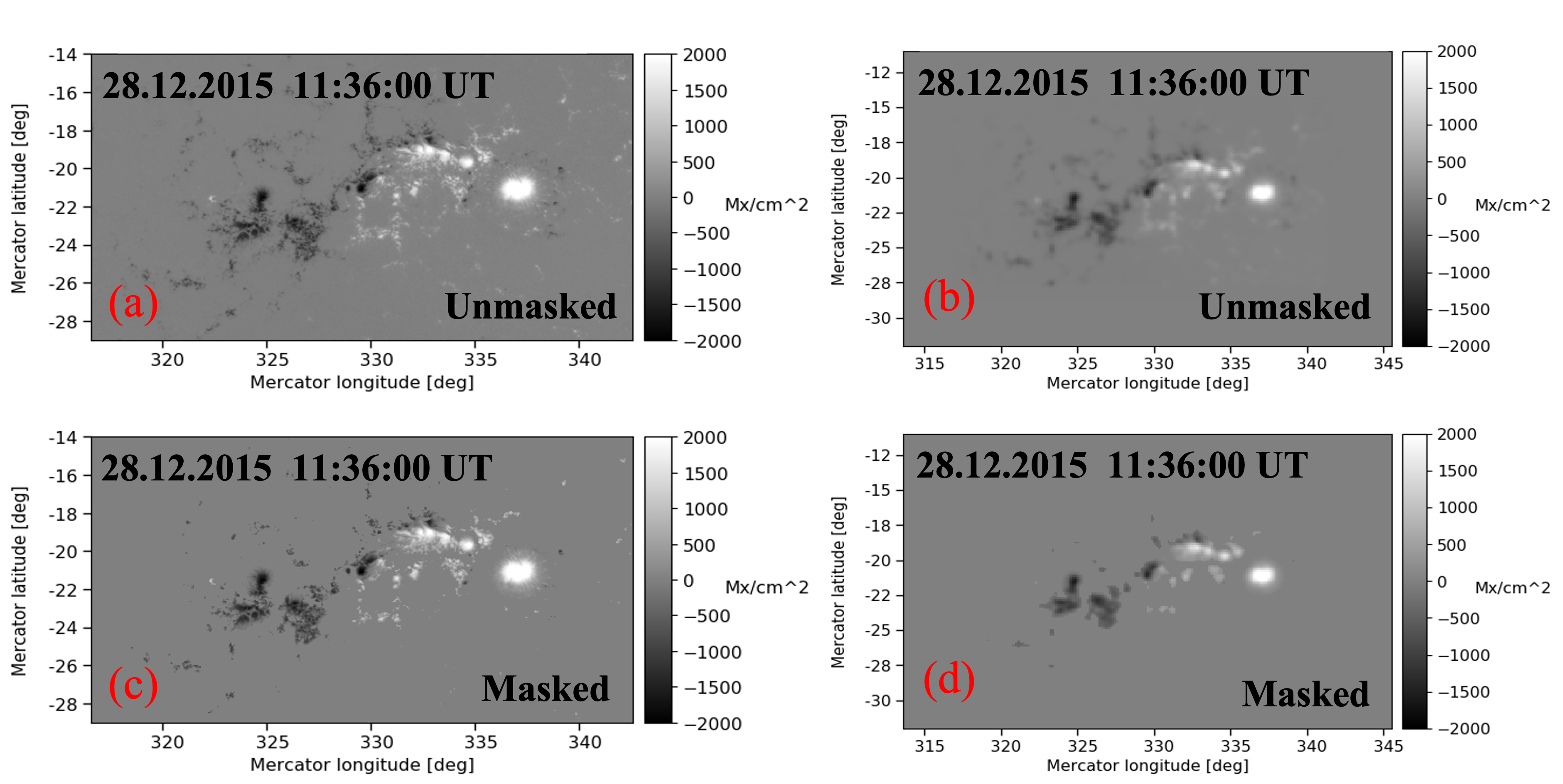}
   \caption{$Bz$ component of magnetogram cutout used in the analysis and simulation taken at 11:36 UT on December 28, 2015 (the time when the eruption was observed): (a) Re-projected cutout where no masking is applied; (c) re-projected cutout where masking is applied with a threshold of $B \leq 250 ~\textrm{Mx} ~\textrm{cm}^{-2}$; (b) simulation ready cutouts produced with ELECTRICIT where no masking is applied; (d) simulation ready cutouts produced with ELECTRICIT where masking is applied with a threshold of $B \leq 250 ~\textrm{Mx} ~\textrm{cm}^{-2}$.}
    \label{Fig:figure1}%
    \end{figure*}

\subsubsection{Electric Field Inversion Methodology}
\label{sec:section2.3.2}

%ELECTRICIT uses the Poloidal-Toroidal Decomposition \citep[PTD,][]{Kazachenko2014} and Faraday's law to determine the  photospheric electric field.
%from the PTD-Doppler-FLCT-Ideal (PDFI) method (here, FLCT is Fourier Local Correlation Tracking) \citep{Kazachenko2014} and 
%\textit{ad hoc} assumptions by \cite{Cheung2012, Cheung2015}.
%and on the DAVE4VM velocity inversion.
%ELECTRICIT uses the Faraday's law to determine the  photospheric electric field \citep{Lumme2017}.
The photospheric electric field $\mathbf{E}$ is decomposed into two components: i) inductive $\mathbf{E}_I$; and ii) non-inductive $- \nabla \psi$, as
\begin{equation}
    \mathbf{E} = \mathbf{E}_I - \nabla\psi.
\end{equation}
The inductive component is determined from 
%is constrained using ELECTRICIT, which uses
Faraday's law directly using the set of photospheric
magnetic field measurements, $\nabla \times \mathbf{E}_I = - \frac{\partial \mathbf{B}}{\partial t}$. 

%\textcolor{red}{inverses faraday's equation and build upon it in simpler form}

In order to determine the non-inductive component which is not constrained by Faraday's law, we employ \textit{ad hoc} assumptions on the form of the divergence of the non-inductive electric field. %They were briefly presented in Section \ref{sec:section1} and 
They are given by \citep{Cheung2012, Cheung2015, Lumme2017, Pomoell2019}:
\begin{equation}
\nabla_h \cdot \mathbf{E}_h = -\nabla_h^2 \psi = 0
 \label{eq:equation2}
\end{equation}
\begin{equation}
\nabla_h \cdot \mathbf{E}_h =
-\nabla_h^2 \psi 
%= - \nabla_h \cdot \mathbf{E}_h 
= - \Omega B_z
 \label{eq:equation3}
\end{equation}
\begin{equation}
\nabla_h \cdot \mathbf{E}_h = -\nabla_h^2 \psi 
%= - \nabla_h^2 \cdot \mathbf{E}_h 
= - U (\nabla \times \mathbf{B}_z)_h
 \label{eq:equation4}
\end{equation}
where, $\Omega$ (units rad/s) and $U$ ( units in Mm/s) are free parameters in the non-inductive electric field, which are assumed to be spatially and temporally constant (see their physical interpretation in \cite{Lumme2017, Pomoell2019}). These assumptions are henceforth referred to as the i) zero- (eq. \ref{eq:equation2}); ii) $\Omega$- (eq. \ref{eq:equation3}) ; and iii) $U$- (eq. \ref{eq:equation4}) 
assumption, respectively. 

\subsubsection{Choice of $U$ and $\Omega$ }
\label{sec:section2.3.3}

For obtaining the values of the constant $U$ and $\Omega$, we optimised them with respect to an independent metric, namely the injection of photospheric magnetic energy as a function of time, $E_M(t)$. %For this, we compared it with the 
As the target, we used the
magnetic energy injection produced by the photospheric electric field %calculated 
%from DAVE4VM using the ideal Ohm's law:
computed using the ideal Ohm's law:
\begin{equation}
    \mathbf{E} = - \mathbf{V}_\mathrm{DAVE4VM} \times \mathbf{B},
    \label{eq:equation8}
\end{equation}
where $\mathbf{V}_\mathrm{DAVE4VM}$ is the photospheric flow velocity vector determined by DAVE4VM \citep{Schuck2008} on the basis of optical flow methods.  %\textbf{We determined $\mathbf{B}$ extracted from unmasked and masked vector magnetograms.}
%to estimate the energy injection from the time series magnetograms. It models image motion with the continuity equation and/or the convection equation to provide all the three components of the velocity fields for any input at any give time. 

The photospheric magnetic energy injection to the upper solar atmosphere is computed by integrating the vertical Poynting flux ($S_z$) over time and over the area ($A$) of the magnetogram \citep{Kazachenko2015}:
\begin{equation}
\begin{aligned}
    E_M(t) = \int_{0}^{t} \,dt' \frac{dE_M}{dt} =
    \int_{0}^{t} \,dt' \int_{}^{} dA ~S_z \\
    =
    \frac{1}{\mu_0} \int_{0}^{t} \,dt' \int_{}^{} dA (\mathbf{E} \times \mathbf{B}) \cdot \mathbf{\hat{z}}
    \label{eq:equation9}
\end{aligned}
\end{equation}
where, $\mu_0$ is the magnetic constant. For a given choice of \textit{ad-hoc} assumption and value for the associated $U$ and $\Omega$, the photospheric electric field can be inverted and the magnetic energy injection determined. Thus, a given reference magnetic energy injection is sufficient to determine the non-inductive component of the driving electric field. The optimised values of the $U$ and $\Omega$ were found with an automated procedure. The procedure consists of a bracketing-like procedure in which the parameter space is discretized and for each value comparing the resulting energy injection to the reference DAVE4VM estimate by computing the root mean square (rms) difference, until a minimum deviation is found. This was done for the data where the magnetic field was masked with a threshold of $< 250$ Mx/cm$^{-2}$ hereinafter called masked data. This procedure (i.e., Eqs. 5-6) was also done for the data without any magnetic field masking, and is hereinafter called unmasked data.
Using this method, the value of optimised $U$ and $\Omega$ was found to be 120 Mm/s and 0.06 rad/s,
for unmasked data, respectively and $200 ~\textrm{Mm/s}$ and $0.13 ~\textrm{rad/s}$, for masked data. The accuracy in the estimates of $U$ and $\Omega$ is the step of increment/decrements used for finding the optimised values, which were $\pm 5 ~\textrm{ms}^{-1}$ and $\pm 0.01 \times 2\pi ~\textrm{day}^{-1}$, respectively for $U$ and $\Omega$ runs. 

The aim is to study the effects that the choice of $U$ and $\Omega$ values have
%optimisation parameters 
on the data-driven coronal simulation. By varying $U$ and $\Omega$ values, we can explore the changes in flux rope 
properties and evolution.
%appearance, evolution and eruption and their properties. 
In order to achieve this, we 
prepared additional electric field data sets using %scaled 
modified values for $U$ and $\Omega$. 
Specifically, we constructed eight additional electric field sets by scaling the optimized
%the simulation input magnetograms for 
$U$ and $\Omega$ values by
the factors 
$0.25, 0.5, 2$ and $4$. 

For the last stage of preparing the data, following our previous work, we smoothed the magnetograms and rebinned. We used $4 \times 4$ pixel rebinning, which resulted in a pixel size of $\approx 1.46 ~\textrm{Mm}$. The electric fields were re-inverted using the smoothed data and with the different values of $\Omega$ and $U$ obtained during optimisation. 
%The possible datagaps in the input magnetogram series were filled with linearly interpolated and smoothed artificial magnetograms, so that the output had no datagaps.

\subsection{TMFM of AR 12473}
 \label{sec:section2.5}
 
The TMFM simulations were carried out from 23:36 UT on December 23, 2015, when the center of AR12473 was approximately at E50  on the disk, until 15:36 UT on January, 02, 2016, at which time the AR had reached $\approx$ W70. The HMI data was prepared as mentioned in the previous sections. The simulation input magnetograms were padded with 25 pixels of zeroes at the boundary to reduce boundary effects due to the finite size of the computational domain. The simulation box height was chosen to be 200 Mm ($\approx 0.29 \rm R_{\odot}$). The boundary conditions of the simulation box were chosen to be open \citep{Pomoell2019}. 

The simulations were ran using the input magnetograms prepared for various different \textit{ad hoc} $U$ and $\Omega$ values. The optimised values of $U$ and $\Omega$ were obtained using the procedure explained in section \ref{sec:section2.3.3}. 
%The simulation input magnetograms were also prepared for 
%$1/4 \times, 1/2 \times, 1 \times, 2 \times \& ~4 \times $ of the optimised values. 
In addition, inputs were also prepared for the electric fields inverted using scaled values for the inversion parameters.
 Table \ref{tab:table1} shows the optimised values for all investigated cases in bold. The errors shown for the optimised parameters in Table \ref{tab:table1} is the step of the increment/decrements used in the optimisation procedure. The simulations were also run using the magnetograms prepared by setting the non-inductive electric field to zero.
 
 For each value of $U$ and $\Omega$, as well as for the case of zero non-inductive electric field, the simulation was run with both  unmasked and masked magnetograms. In the latter case we used the masking threshold $B = 250 ~\textrm{Mx} ~\textrm{cm}^{-2}$, i.e. all pixels for which $B \leq 250 ~\textrm{Mx} ~\textrm{cm}^{-2}$ were assigned to $0$. As an example, Figure \ref{Fig:figure1} shows the excerpts of vertical component ($Bz$) of raw re-projected magnetograms cutouts from HMI during the eruption (11:36 UT on December 28, 2015).  The left panel contains both unmasked and masked $Bz$ component of magnetograms (threshold: $B \leq 250  ~\textrm{Mx} ~\textrm{cm}^{-2}$). 
The right panel shows the $Bz$ component of simulation ready cutouts (re-binned and padded) produced with ELECTRICIT during the same time for one of the simulations %(parameter: U) 
for both the masked and unmasked case. 

    \begin{figure*}
   \centering
    \includegraphics[width=1\textwidth,clip=]{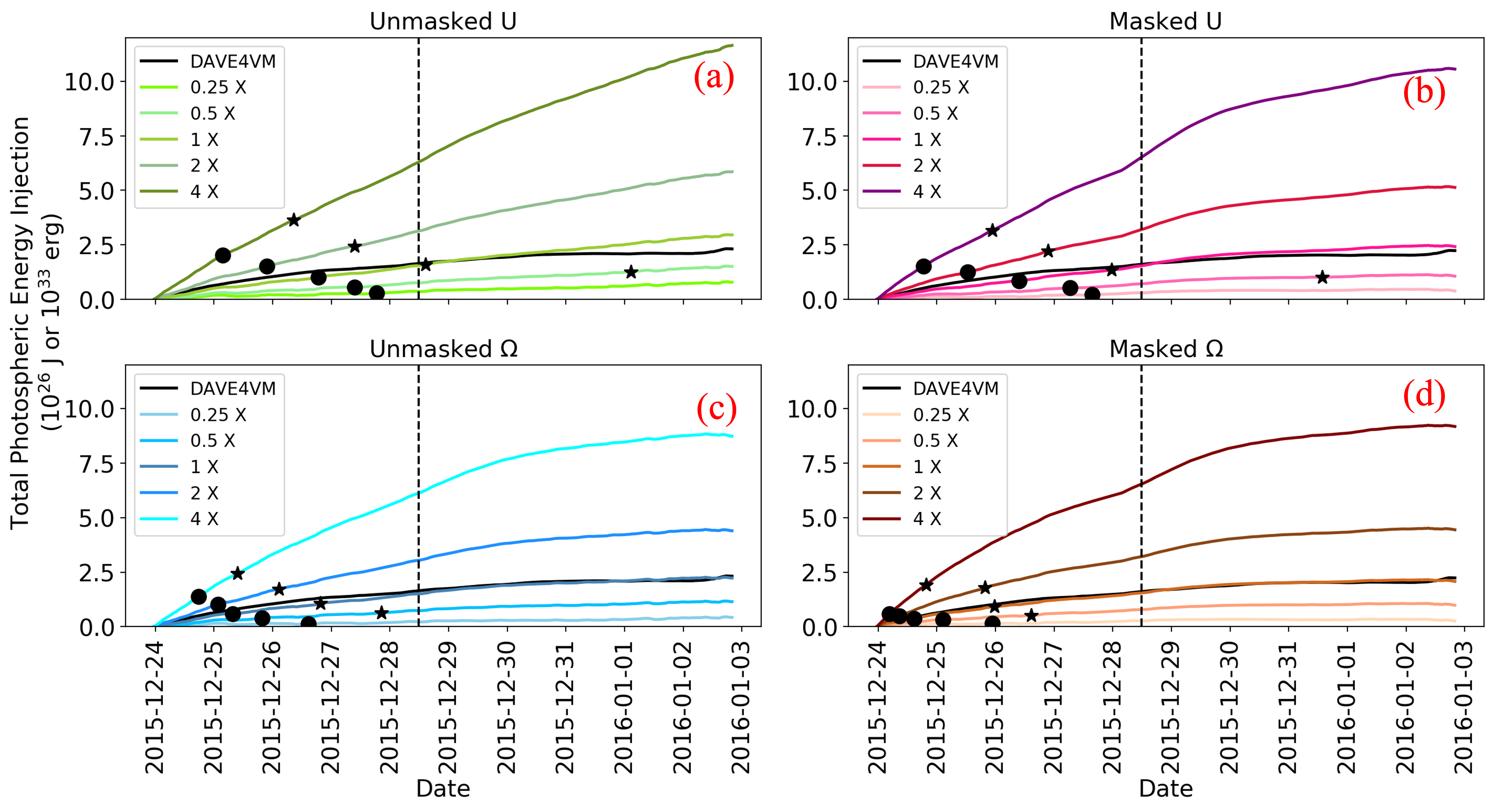}
   \caption{Temporal evolution of total photospheric magnetic energy injection for AR12473 from 23:36 UT on December 23, 2015 to 08:36 UT on January 02, 2016. The black dot and black star represent the flux rope formation time and the time when the flux rope reaches $\approx 100  ~\textrm{Mx}$ for various values of $U$ and $\Omega$: a) the unmasked $U-$ assumption; (b) the masked $U-$ assumption; (c) the unmasked $\Omega-$ assumption; and (d) the masked $\Omega-$ assumption. The free parameters in the non-inductive electric field values vary from $0.25 \times $ to $4 \times$ of the optimised values. The DAVE4VM reference is also shown in solid black color. The vertical dashed line is the M1.9 class flare peak time on Dec 28, 2015.}
              \label{Fig:figure2}%
    \end{figure*}

\section{Results}
\label{sec:section3}

In this section, we present the simulation results for AR 12473 using different values of the $U$ and $\Omega$ used in the photospheric electric field inversion (Table 1) and investigate their effects on the A . 

\subsection{Photospheric Energy and Relative Helicity Injection}
\label{sec:section3.1}

First, we investigated the photospheric energy injection, computed using eq. \ref{eq:equation9}, for various values of the parameters $U$ and $\Omega$. The results were compared with the energy injection obtained from the DAVE4VM optical flow-based electric
field estimate that was also used to find the optimal values for $U$ and $\Omega$ (see Section \ref{sec:section2.3.3} and Table 1).

    \begin{figure*}
   \centering
        \includegraphics[width=1\textwidth,clip=]{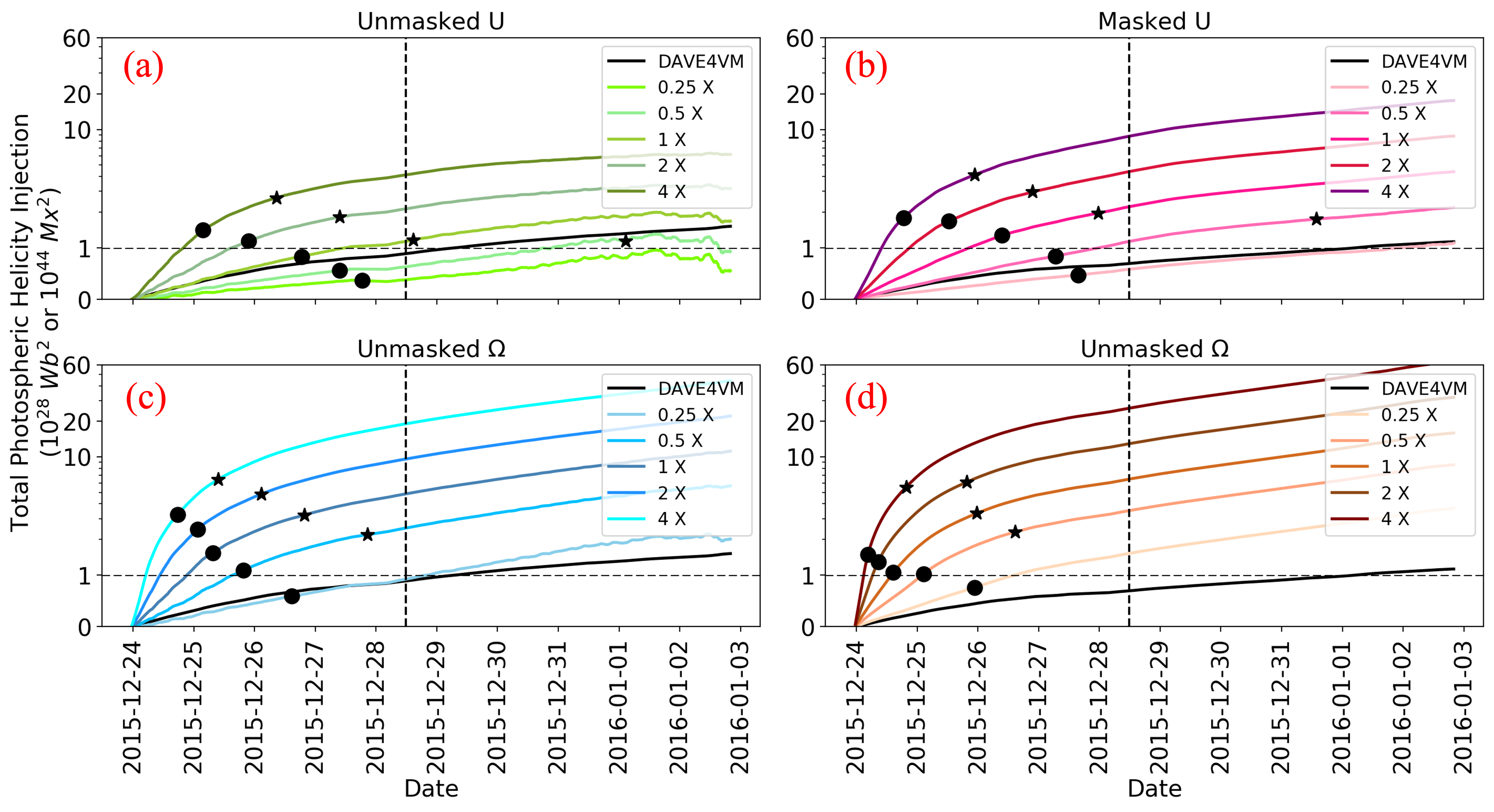}
   \caption{Temporal evolution of total photospheric helicity injection for AR12473 from 23:36 UT on December 23, 2015 to 08:36 UT on January 02, 2016. The black dot and black star represent the flux rope formation time and the time when the flux rope reaches $\approx 100 ~\textrm{Mm}$ for various values of $U$ and $\Omega$: a) the unmasked $U-$ assumption; (b) the masked $U-$ assumption; (c) the unmasked $\Omega-$ assumption; and (d) the masked $\Omega-$ assumption. The free parameters in the non-inductive electric field values vary from $0.25 \times $ to $4 \times$ of the optimised values. The DAVE4VM reference is also shown in solid black color. The vertical dashed line is the M1.9 class flare peak time on Dec 28, 2015. }
              \label{Fig:figure3}%
    \end{figure*}

Figure \ref{Fig:figure2}  shows the energy injection for all the $U$ and $\Omega$ values as summarized in Table \ref{tab:table1}. As noted in section \ref{sec:section2.3.1}, during the optimisation process the full-resolution and unsmoothed vector magnetograms were used. The energy injection closely matches with the DAVE4VM reference curve (black) for the optimised values throughout the displayed time for all four optimized datasets. At the time of the observed eruption (i.e, 28 December, 2015 at 11:36 UT, vertical dashed line) for $0.25 \times$ the optimised value the energy injection is about $\approx 80 \pm 2\%$ less than the DAVE4VM value, and for $0.5 \times$ the optimised value about 54\% less. The percentages do not vary significantly between the $U$ and $\Omega$ assumptions, nor with the masked and unmasked cases. We note that the optimisation procedure minimizes the RMS difference for the entire considered time-interval and not for a specific instance of time. It is also important to note that the value of the optimized $U$ and $\Omega$ values differ for the unmasked and masked cases. For $2 \times$ of the optimised values the energy injection is about 94\% larger and for $4 \times$ of the optimised values about 290\% larger, i.e., the increase is approximately linear with the increase in the parameter value. The overestimation for $2 \times$ and $4 \times$ optimised values increases at later times. 
%The 'black dot' and 'black star' represent the start time and the time when the flux rope reaches $\approx 100 ~Mm$ for various values of $U$ and $\Omega$
The dot and star markers in the figures 
represent 
for each simulation the time of the formation of the flux rope and the time
at which the flux rope reaches 100 Mm, respectively, and is detailed in section \ref{sec:section3.3}.

%Table \ref{tab:table1} also lists the time duration taken by the flux rope to reach the above mentioned height for various simulations.

Figure \ref{Fig:figure3} shows the temporal evolution of the photospheric injection of relative helicity for the same electric field inversions as provided in Figure \ref{Fig:figure2}, together with the corresponding DAVE4VM result. The total photospheric helicity injection is larger than the estimate from DAVE4VM for all $\Omega$ runs and masked $U$ runs \citep[Eq. 18 in][]{Pomoell2019}. The injected helicities for $0.5 \times$ and in particular so for $0.25 \times$ the optimised values is however very close ($\pm 20 \%$)
to that obtained from DAVE4VM (see figure \ref{Fig:figure3}c). In addition, for the unmasked $U$ case for $0.25 \times$ and $0.5 \times$ the optimised values were lower than the DAVE4VM estimate and the energy injection for the optimised $U$ matched also relatively closely with DAVE4VM. With $2 \times$ and  $4 \times$ the optimised values led to a drastic overestimation of the helicity injection when compared to DAVE4VM. We also noticed that for the optimised datasets, the helicity estimates were $2-3 \times$ higher in $\Omega$ runs than $U$ runs. This effect was seen during the flux rope evolution (discussed in detail in section \ref{sec:section3.3}). Similar to Figure \ref{Fig:figure2}, we did not notice much relative variation in the helicity injection between masked and unmasked $\Omega$ data sets. 
\subsection{Formation and Evolution of the Flux Rope}
\label{sec:section3.3}

\subsubsection{Flux Rope Identification}
\label{sec:section3.3.1}

For the identification of the flux rope in the simulation, we used the twist number $T_w$, which quantifies the winding of two infinitesimally close field lines about each other. It is defined as, $T_w = \frac{1}{4 \pi} \int ds \, \mu_0 J_{\parallel}/B$ \citep{Berger2003}, where $J_{\parallel}$ is the current density parallel to the magnetic field and $ds$ is the arc length increment along the magnetic field line.
To facilitate locating the flux rope using the twist number, we compute the value of $T_w$ in a plane that is approximately perpendicular to the local axis of the flux rope. Thus, we placed the plane at $y = 0$, i.e. vertical to the XZ plane. For all simulation runs using a non-zero non-inductive electric field component we were able to identify a coherent region of $T_w \geq 1.5$ which rose in height through the simulation domain with increasing time. We consider the volume of space filled by the field lines passing through the coherent $T_w$ region as the flux rope (FR). With the FR defined, we tracked the evolution of the structure until the height of 200 Mm (upper edge of the simulation domain) and determined various FR parameters, such as, accumulated helicity, axial flux and axial magnetic field. For the cases where the non-inductive component of the driving photospheric electric field was set to zero (last two rows of Table 1), a flux rope as defined above did not form during the simulation time.
%The magnetic field lines follow the path of the flux rope for which the $T_w$ was greater than 1.5. In the simulation, we found the coherent region with $T_w \geq 1.5$, which was seen progressing in height with time. 

\begin{figure*}
   \centering
    \includegraphics[width=1\textwidth,clip=]{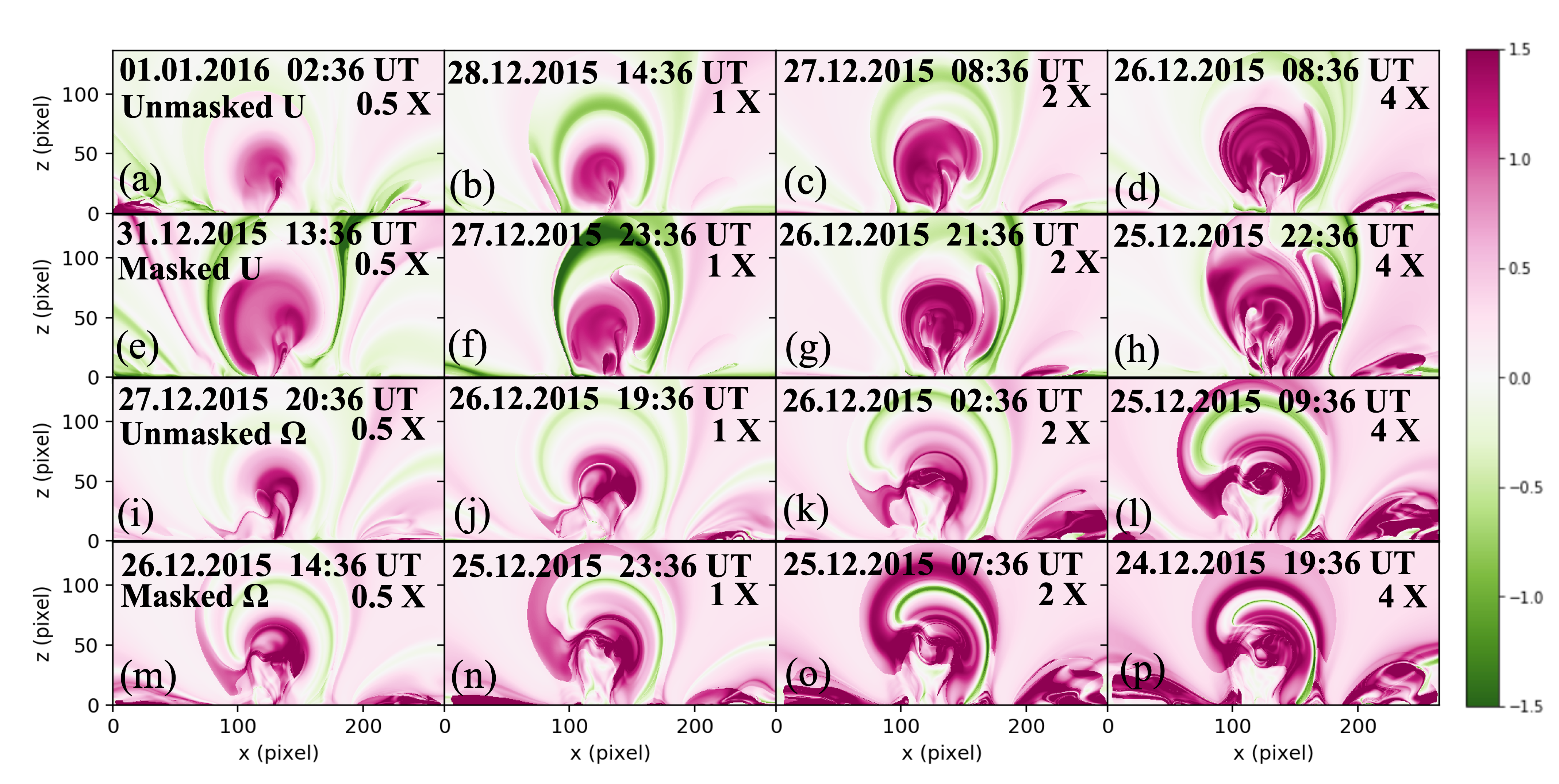}
   \caption{The figure shows the snapshots of the twist number $T_w$ maps (green-purple) for different TMFM runs.
   }
 \label{Fig:figure6}%
    \end{figure*}

\begin{figure*}
   \centering
    \includegraphics[width=1\textwidth,clip=]{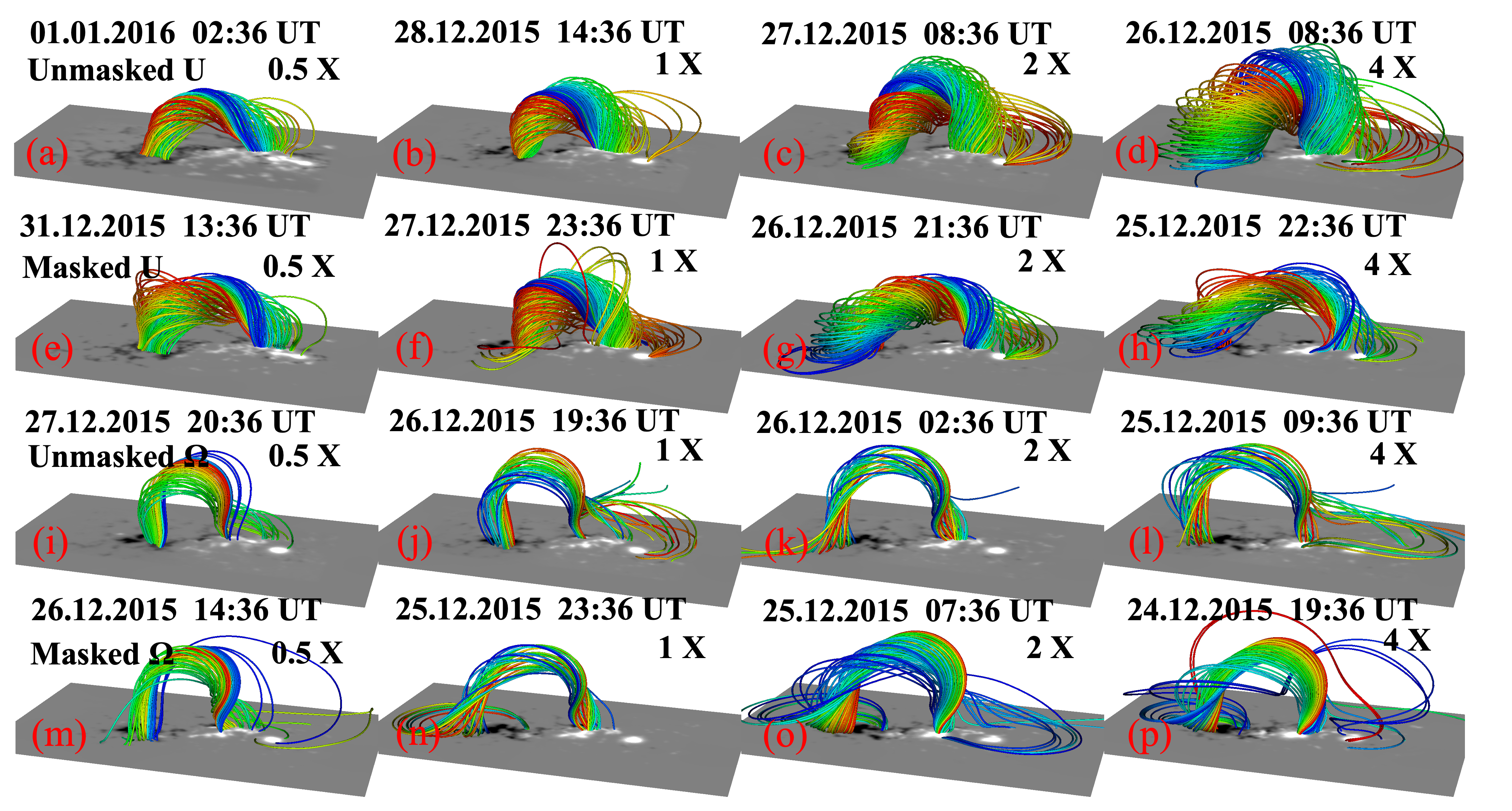}
   \caption{Excerpts from different simulations showing the time when the flux rope reaches the mid height in the simulation domain.
   %he gray disc shows the area for which the total axial flux was calculated. The blue vertical lines show axis along which the axial magnetic field was estimated. 
   }
              \label{Fig:figure7}%
    \end{figure*}

%We tracked the FR as it rose and expanded in the box (height 200 Mm).
Snapshots of the $T_w$ planes at the times when the apex of the identified FR reached $\approx 100 ~\textrm{Mm}$ are given in  Figure \ref{Fig:figure6}, while a visualization of selected magnetic field lines through the $T_w \geq 1.5$ region for each corresponding time is shown in Figure \ref{Fig:figure7}. Figure \ref{Fig:figure6} clearly demonstrates that all simulations, irrespective of optimisation value, generated a consistently positively twisted flux rope. 
The figure also shows that the twist values are higher for higher values of $\Omega$ and $U$. Although the overall appearance of the FRs are very similar between $U$ and $\Omega$ runs the twist structure and visualized magnetic field lines of the FRs appear more coherent for the $U$ runs. In particular for some of the $\Omega$ runs, there are some field lines that exit the simulation domain through the lateral boundaries that are associated with the FR.
The twist maps also reveal a significant negative region above the apex of the positively twisted FR. %With time, the lower boundary of $\Omega$ runs becomes busy (see Figure \ref{Fig:figure6} (lower panel). 

 \subsubsection{Magnetic Energy and Relative Helicity in the Corona}
\label{sec:section3.2}

Next, we quantified the effect of the different
driving photospheric electric fields on the 
evolution of the coronal magnetic field in the TMFM simulation.
%which was further processed after optimisation (section \ref{sec:section3.1}) as,
We computed the total magnetic energy ($\varepsilon_M$) and free magnetic energy ($\varepsilon_{\text{free}}$) in the entire coronal simulation volume as   
\begin{equation}
    \varepsilon_M = \frac{1}{2\mu_0} \int dV \, B^2
\end{equation}
\begin{equation}
    \varepsilon_{free} = \frac{1}{2\mu_0} \int dV \, (B^2 - B_p^2)
\end{equation}
where $\mathbf{B}_p$ is the associated potential field. 
%
%The results are shown in Figure 4 for the energy and in Figure 5 for the relative helicity. Note that, these volume metrics were calculated for the entire simulation box and not only for the region close to the flux rope. 
%shown in figure \ref{Fig:figure7}.
%It is important to note that TMFM simulation was driven by 

\begin{figure*}
   \centering
    \includegraphics[width=1\textwidth,clip=]{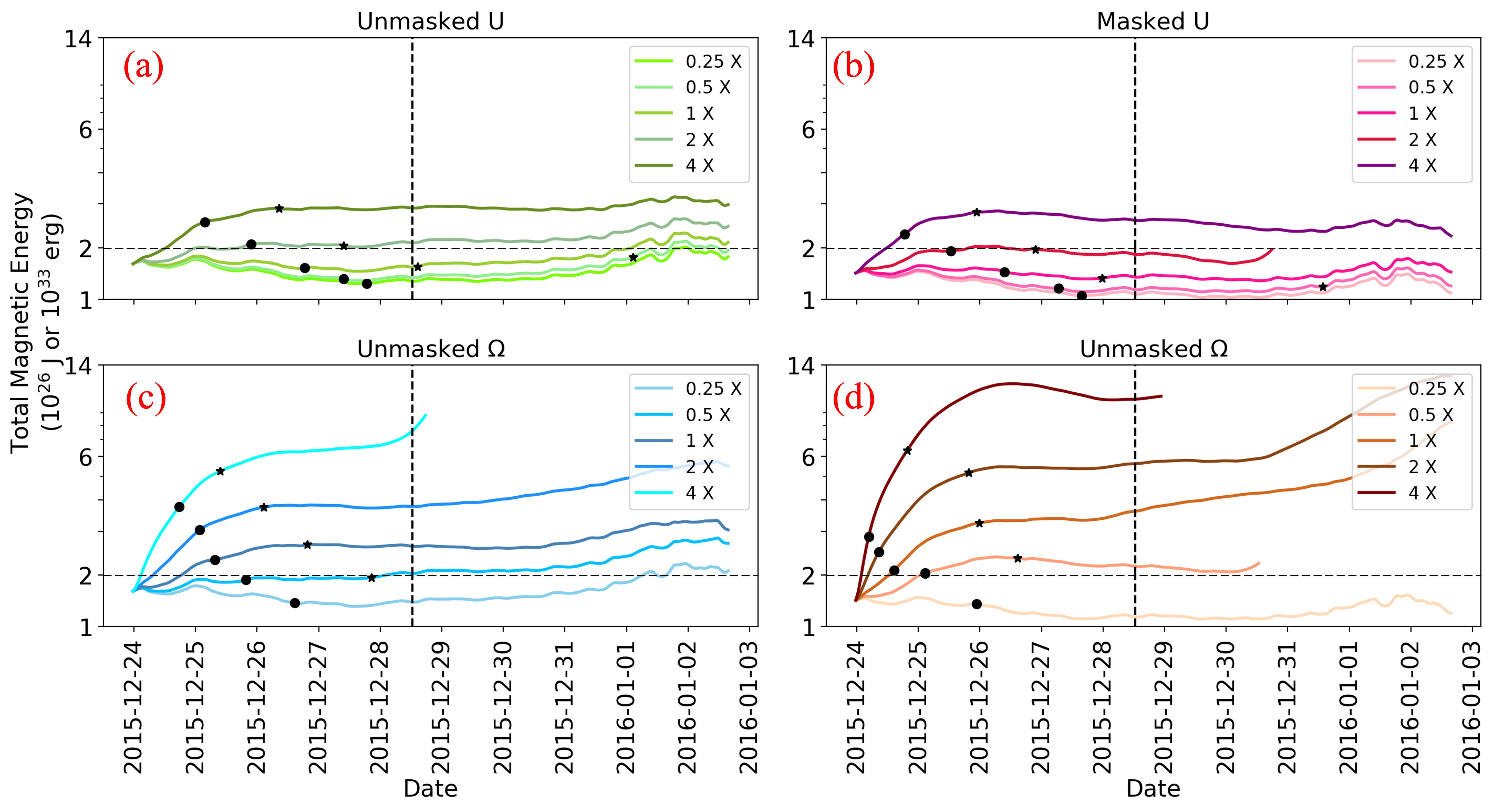}
    \includegraphics[width=1\textwidth,clip=]{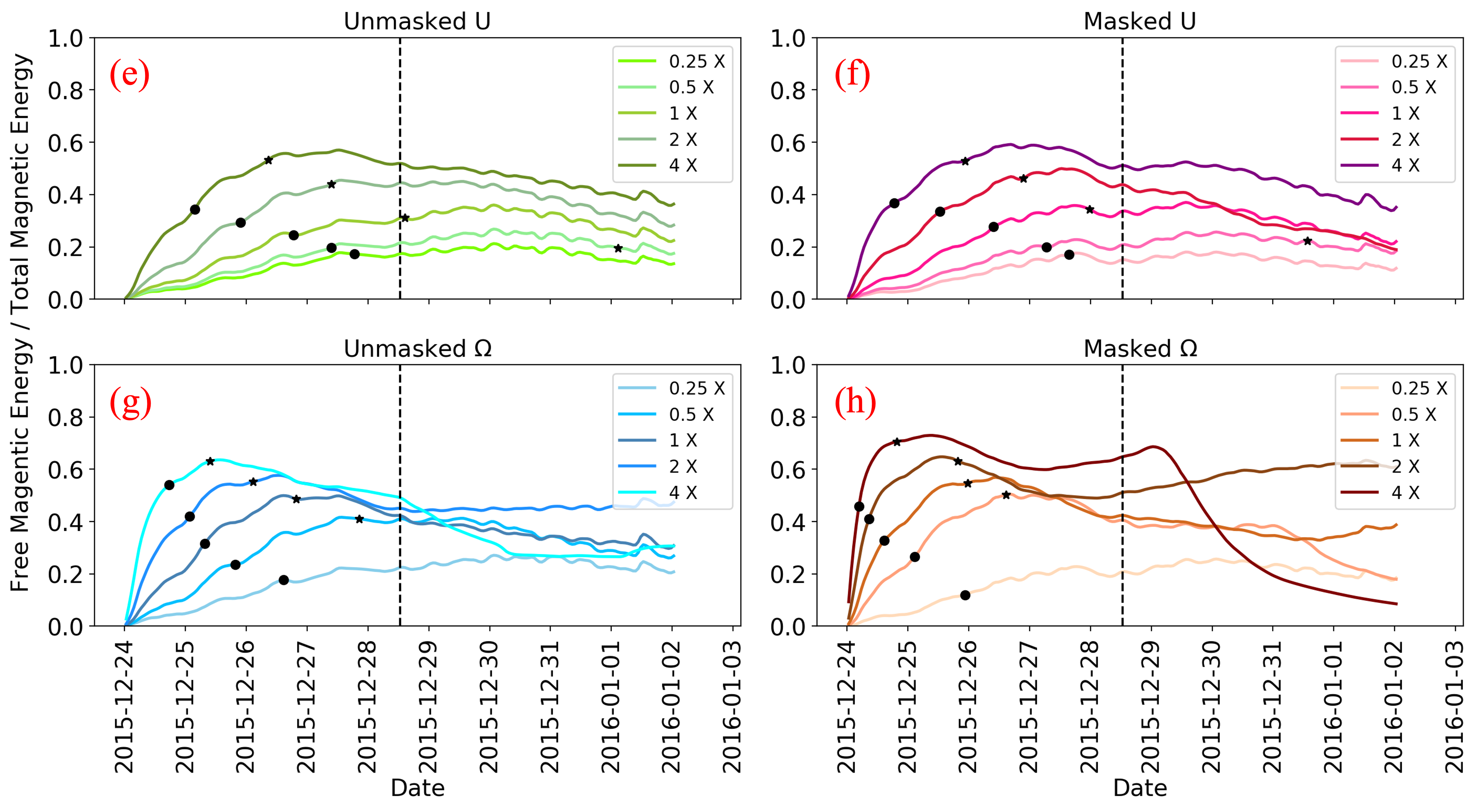}
   \caption{Top Panel: Temporal evolution of total magnetic energy for AR12473 from 23:36 UT on December 23, 2015 to till the time the flux rope reached $\approx 100 ~\textrm{Mm}$ for various values of $U$ and $\Omega$ using the processed simulation data: a) the unmasked $U-$ assumption; (b) the masked $U-$ assumption; (c) the unmasked $\Omega-$ assumption; and (d) the masked $\Omega-$ assumption.  The black dot and black star  represent the flux rope formation time and the time when the flux rope reaches $\approx 100 ~\textrm{Mm}$ for various values of $U$ and $\Omega$.
   The free parameters in the non-inductive electric field values vary from $0.25 \times $ to $4 \times$ of the optimised values.  
   Bottom Panel: Same as above but for the ratio of the free to the total magnetic energy. The vertical dashed line is the M1.9 class flare peak time on Dec 28, 2015.}
              \label{Fig:figure4}%
    \end{figure*}

Figure \ref{Fig:figure4} shows that the evolution of total energy varies significantly depending on the driving electric field. 
For unmasked and masked $0.25 \times$, $0.5 \times$ and $1 \times$ optimised $U$ runs, the total energy first decreases reaching minimum around the time of flux rope formation (section \ref{sec:section3.3}) and slowly increases. For $2 \times$ and $4 \times$ optimised $U$ the energy initially shows a sharp increase before reaching a plateau. 
%For $2 \times$ case this occurs when the flux rope forms and for the  $4 \times$ case when the flux rope reaches 100 mM. 
For $\Omega$ runs, the changes in total energy for $0.25 \times$ and $0.5 \times$ optimised value runs are more subtle, while the $4 \times$ optimised run does not reach the plateau during the tracked flux rope evolution (see section \ref{sec:section3.3} for details).

The magnetic energy followed a very similar pattern for all of the runs with lower values of optimised $\Omega$ and $U$. For $U$ $0.25 \times$ and $0.5 \times$ runs, the magnetic energy reaches minimum during the same time, which indicates the flux ropes would have formed at very close by times. These volume metrics shows similar pattern for all the $U$ values. However, for $\Omega$ runs, the total magnetic energy for higher values becomes very high after the flux rope reaches at $\approx 100 ~\textrm{Mm}$.
 The $4 \times$ runs never reached minimum, which indicates that the flux rope was formed at the very beginning of the simulation (refer Table \ref{tab:table1} for details).
The $\varepsilon_{M}$ peaks at $\approx$ 23:36 UT on 25 Dec, for all $U$ runs (except $4 \times$), for both masked and unmasked values. It reaches the minima at $\approx$ 18:36 UT on 27 Dec (except $4 \times$).
Contrary to this, for different values of $\Omega$, $\varepsilon_{M}$ always increased between flux rope formation until it reached the middle of the simulation box.
%\textcolor{red}{Ask Jens about Flux rope formation and flux cancellation w.r.t free energy? It never decreases. }

The  ratio of the free magnetic energy to total magnetic energy ($\varepsilon_{\textrm{free}} / \varepsilon_M$) for different values of $U$  and $\Omega$ are shown in Figure \ref{Fig:figure4} (lower panel). For higher values of $U$, $\varepsilon_{\textrm{free}} / \varepsilon_M$ increases during the FR emergence until it reaches the height $\approx 100 ~Mm$. For lower values of $U$, $\varepsilon_{\textrm{free}} / \varepsilon_M$ value remains the same during the time of FR emergence until it reaches the height $\approx 100 ~Mm$.
Contrary to this, $\varepsilon_{\textrm{free}} / \varepsilon_M$ increases for all the values of $\Omega$, irrespective of mask effect. 
%\textcolor{red}{Ask Jens if free energy is converted in K.E during the eruption?  }

The relative helicity ($H_R$) and the current-carrying helicity ($H_j$) were calculated in the simulation  as in \cite{Berger2003}. 
\begin{equation}
    H_R = \int dV \, (\mathbf{A}+\mathbf{A}_p) \cdot (\mathbf{B}-\mathbf{B}_p) = H_j+2H_{pj}
\end{equation}
\begin{equation}
    H_j = \int dV \, (\mathbf{A}-\mathbf{A}_p) \cdot (\mathbf{B}-\mathbf{B}_p) 
\end{equation}
\begin{equation}
    H_{pj} = \int dV \, \mathbf{A}_p \cdot (\mathbf{B}-\mathbf{B}_p)
\end{equation}
where, $\mathbf{A}$ and $\mathbf{A}_p$  are the vector potential of magnetic field $\mathbf{B}$ and potential magnetic field $\mathbf{B}_p$, respectively and $H_{\textrm{pj}}$ is the mutual helicity between $H_j$ and $H_R$. 

\begin{figure*}
   \centering
    \includegraphics[width=1\textwidth,clip=]{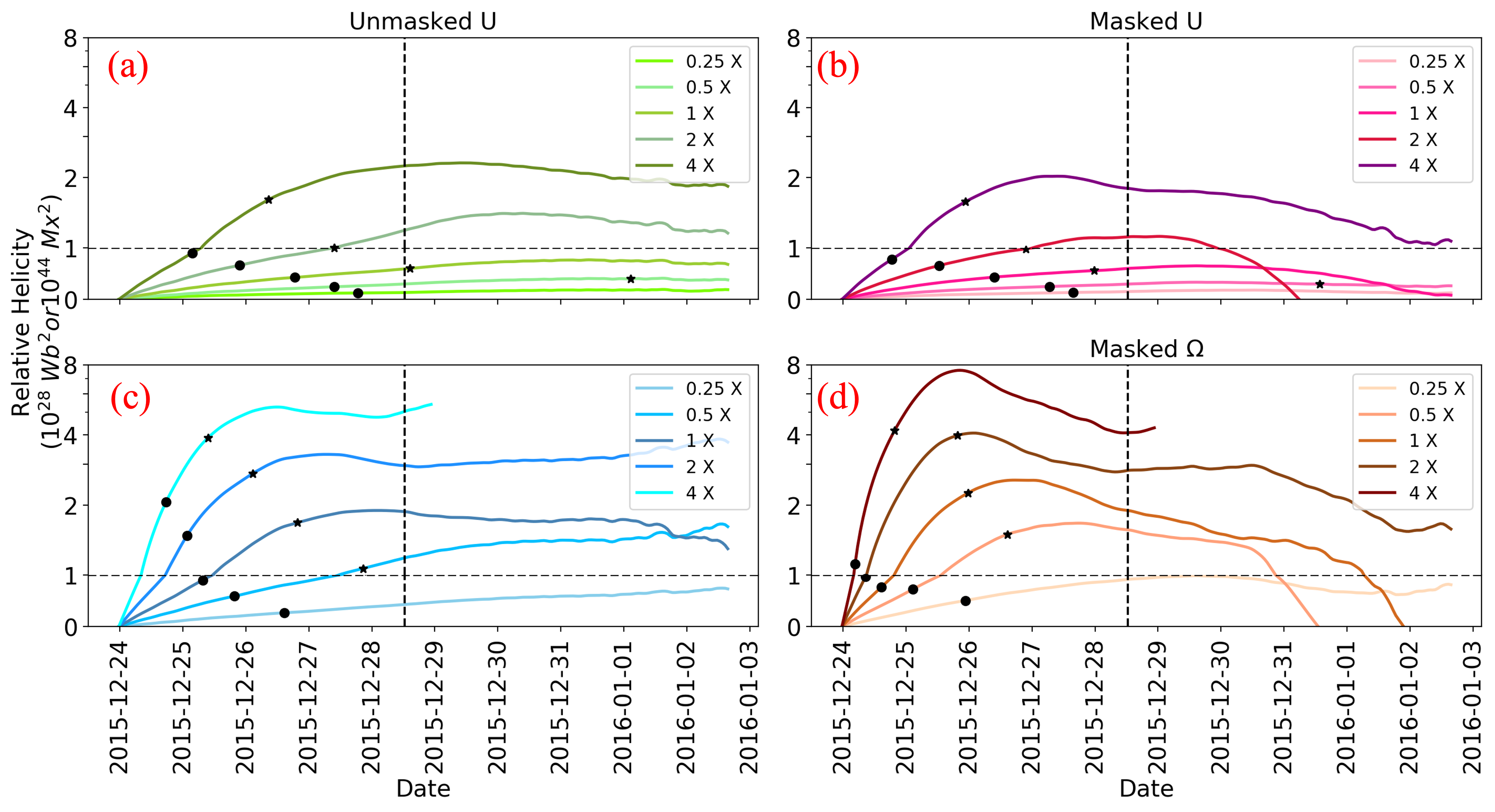}
    \includegraphics[width=1\textwidth,clip=]{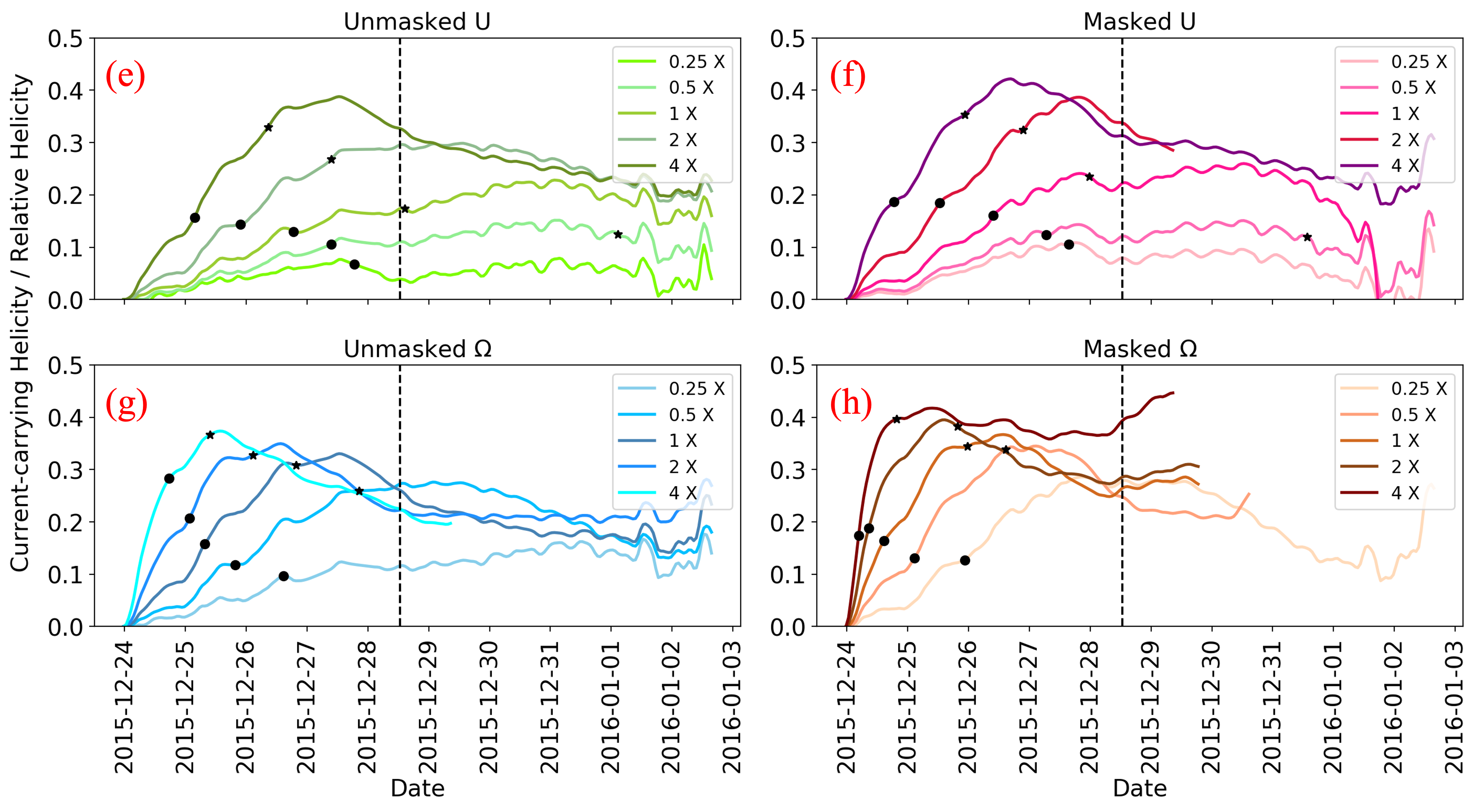}
   \caption{Top Panel: Temporal evolution of relative helicity for AR12473 from 23:36 UT on December 23, 2015 to till the time the flux rope reached $\approx 100 ~\textrm{Mm}$ for various values of $U$ and $\Omega$ using the processed simulation data: a) the unmasked $U-$ assumption; (b) the masked $U-$ assumption; (c) the unmasked $\Omega-$ assumption; and (d) the masked $\Omega-$ assumption. 
   The black dot and black star represent the flux rope formation time and the time when the flux rope reaches $\approx 100 ~\textrm{Mm}$ for various values of $U$ and $\Omega$. The free parameters in the non-inductive electric field values vary from $0.25 \times $ to $4 \times$ of the optimised values.  
   Bottom Panel: Same as above but for the ratio of the current-carrying to the relative helicity. The vertical dashed line is the M1.9 class flare peak time on Dec 28, 2015.}
              \label{Fig:figure5}%
    \end{figure*}

Figure \ref{Fig:figure5} (upper panel) shows the temporal evolution of the relative helicity for various values of $U$  and $\Omega$.  We found that $H_R$ always increased for all the runs between FR formation time until it reached the middle of the simulation domain, irrespective of the optimisation parameter values.  
%There was significant decrease in $H_R$  once the FR reached height $\approx 100 ~Mm$ for .
The ratio of the current-carrying helicity to the relative helicity ($H_j / H_R$) is shown in Figure \ref{Fig:figure5} (bottom panel). 
It can be seen that $H_j / H_R$ shows a similar rise and then decline like $\varepsilon_{\textrm{free}} / \varepsilon_M$ (see Figure \ref{Fig:figure4}, bottom panel). However for all the $U$  and $\Omega$ values, the $H_j / H_R$ ratio had increased between the time of FR formation and till it reaches the height $\approx 100 ~Mm$.
It was previously reported by \cite{Zuccarello2018} that for their torus unstable flux ropes, the $H_j / H_R$ has a threshold of $\approx 0.29 \pm 0.01$, but that it likely is not universal. We found the $H_j / H_R$ value close to this threshold only when the optimisation parameters were close to the optimised values.  Otherwise, the  $H_j / H_R$ value varied between $\approx 10-40 \%$ from the threshold value found by \cite{Zuccarello2018}. This means that in the TMFM runs the flux rope can rise through the simulation domain for a relatively large range of $H_j / H_R$ values. We note that \cite{Rice2022} found using two-dimensional magnetofrictional simulations that this eruptivity index has only a weak predictive skill, and depends on the orientation of the overlying magnetic field with that of the flux rope. 
%We found that it only hold true if the values of optimisation parameters are close to the optimised values. 

\subsubsection{Flux Rope Parameters}
\label{sec:section3.3.2}

\begin{figure}
   \centering
    \includegraphics[width=0.5\textwidth,clip=]{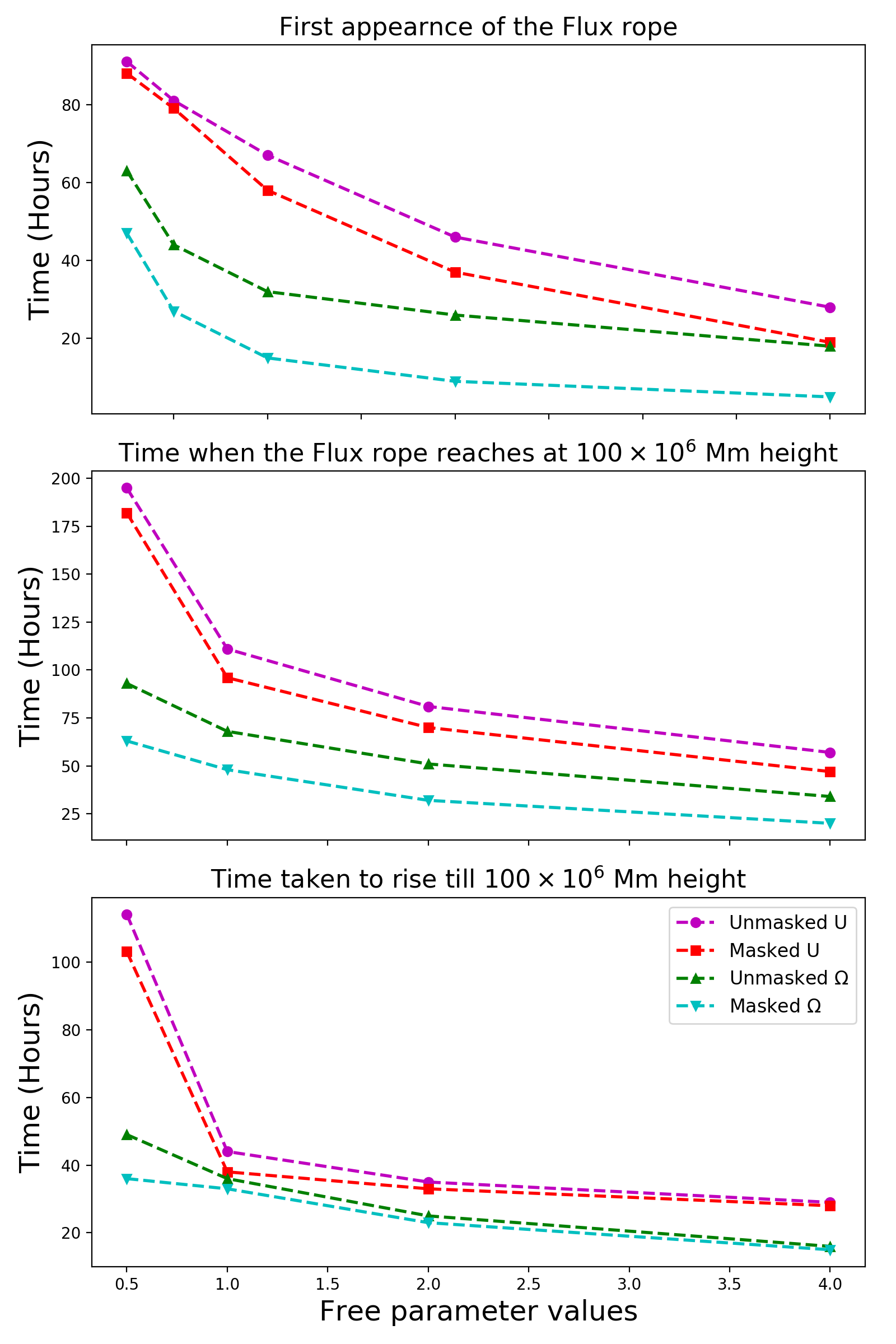}
   \caption{ Time (in hours) for the flux rope: (a) to first appear in the simulation domain; (b) to reach the middle height of the simulation domain; and (c) from first appearance to reach the middle of the simulation domain.
   \textcolor{red}{}}
              \label{Fig:figure8}%
    \end{figure}

\begin{figure}
   \centering
    \includegraphics[width=0.6\textwidth,clip=]{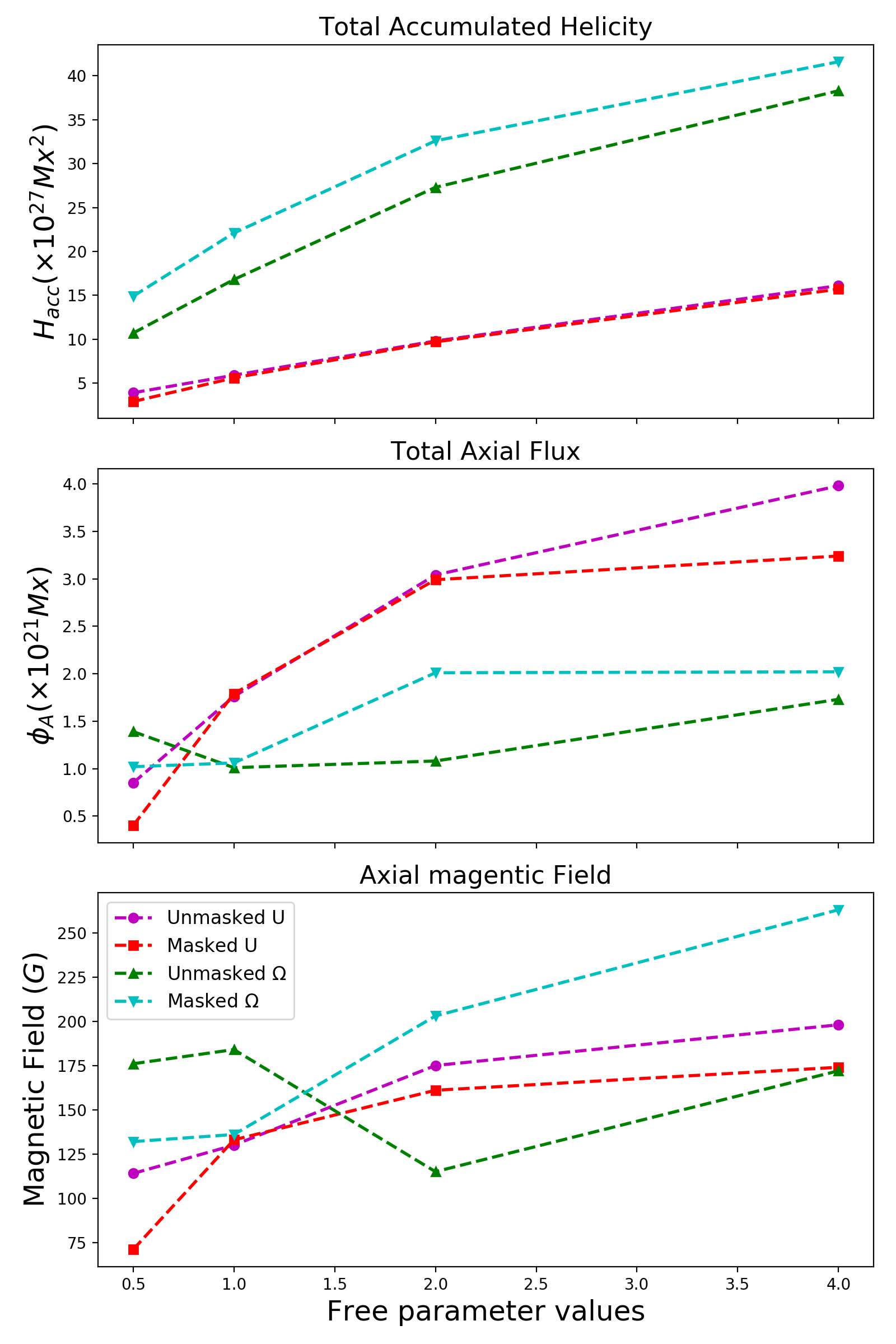}
   \caption{Various flux rope properties when the flux rope apex had reached the altitude of $\approx 100 ~\textrm{Mm}$: (a) the helicity accumulated from the time the flux rope formed to the observed eruption time; (b) the axial flux ; (c) the axial magnetic field magnitude; (d) the flux rope radius; and (e) the peak current-carrying helicity.}
              \label{Fig:figure9}%
    \end{figure}

The following properties were investigated for the identified FRs and they are listed in Table \ref{tab:table1}:  i) flux rope formation time, ii) the time when the FR apex reached $\approx 100 ~\textrm{Mm}$ height; iii) the rise time of the FR to $\approx 100 ~\textrm{Mm}$; iv) accumulated helicity; v) total axial flux; vi) axial magnetic field; and vii) flux rope radius. In addition, Figure \ref{Fig:figure8} and Figure \ref{Fig:figure9} shows flux rope appearance and various flux rope parameters for different values of the optimisation parameters, respectively. 
The total accumulated relative helicity (see Eq. \ref{eq:equation9}) was calculated for the complete simulation box. The axial flux was computed as $\Phi_A = \int_{A} \mathbf{B} \cdot \,d\mathbf{A}\ $, where A is the area of integration in a plane normal to the flux rope axis. The flux rope radius was used to determine the extent of the flux rope (see Figure \ref{Fig:figure7}). The flux rope radius was determined at the apex by fitting a vertical line to its extend in Z-direction. 
%the flux rope apex radius was consider. 

Table \ref{tab:table1} and Figure \ref{Fig:figure9} shows that the FR forms earlier for $\Omega$-runs than for the $U$-runs. This is the case for all investigated values of $\Omega$ and $U$, although the difference decreases with the increasing value. The formation time also clearly decreases with increasing value of the optimised parameter. For example, in the simulation runs performed using $0.25 \times$ the optimized value the FR forms 15 - 24 hours later than in the corresponding optimised value run. The difference in the formation time is smaller for the $\Omega$-runs than for the $U$-runs and for the masked runs than for the un-masked runs. For the values larger than the optimized values the FRs form earlier than for the optimised value runs. In the runs conducted with $4 \times$ the optimised value the FR forms 10 - 39 hours earlier than in the corresponding optimised value run, the largest difference being for the unmasked $U$-run.

Approximately similar trends are also found for the times when the FR reaches $\approx 100 ~\textrm{Mm}$ altitude and the rise time of the FR. This height is reached earlier for the $\Omega$ runs than for the $U$ runs. For all the runs made with $0.25 \times$ the optimised value the evolution of the FR was so slow that its apex did not reach the $\approx 100 ~\textrm{Mm}$ altitude during the simulation time. It is also noteworthy that the rise of the FR is very slow for  $0.5 \times$ the optimised value, but the difference in the rise time is not that large between the runs made with the optimised value, $2 \times$ and $4 \times$ the optimized value.

For the $U$-runs (both masked and unmasked) the total current carrying helicity, total axial flux, axial magnetic field and current carrying helicity increases monotonically with the increasing value of $U$.
For $\Omega$-runs in turn some variations are detected. This could be due to less coherent FR structure in $\Omega$-runs as discussed previously. The FR radius increases monotonically only for the unmasked $U$-run. For masked runs it is lower for $2 \times$ and $4 \times$ optimized $U$-runs than for the optimized $U$-run.

%The flux rope parameters values vary uniformly for $U$ runs when the $U$ values are increased.  However, for $\Omega$ runs, a similar pattern is observed except for $2 \times \Oemga$ runs. This could be because of the very fast eruption in case of $\Omega$ runs.  

\begin{sidewaystable*}
\centering
\caption{Summary of the key results for different U and $\Omega$ runs. Bold $U/\Omega$ indicate the value that was optimised with DAVE4VM energy injection. $H_{acc}$ is the helicity accumulated from the time the flux rope formed to the observed eruption time. $phi_{A}$ is the axial flux and $B_A$ the axial magnetic field magnitude, both calculated when the flux rope apex had reached the altitude of $\approx 100 ~\textrm{Mm}$.}
\label{tab:table1}
\begin{tabular}{lccccccccccc}
\hline
\hline
Event \&         & Mask & FR forms & FR rises & FR Formation & FR Time & Time to & $H_{acc}$ & $\phi_{A}$ & $B_A$ & FR \\
$U/\Omega$-value &      & (Y/N)    & (Y/N)  & time & at 1.15 $R_{\odot}$ & Rise & $\times 10^{27}$ & $\times 10^{21}$  & & Radius
\\
                 &      &         &         & (UT) & (UT) & Hours & {[}Mx$^2${]} & {[}Mx{]}   & {[}G{]} & Mx
\\
\hline
\hline
$\mathrm{U}=30$     & 0  & Y & Y & 27/12/2015 18:36:00 & XX                  &  XX &   XX &  XX &  XX & XX \\
$\mathrm{U}=60$     & 0  & Y & Y & 27/12/2015 08:36:00 & 01/01/2016 02:36:00 & 114 &  3.9 & 0.85 & 114 & 24 \\
$\textbf{U = 120}$  & \textbf{0}  & \textbf{Y} & \textbf{Y} & \textbf{26/12/2015 18:36:00} & \textbf{28/12/2015 14:36:00} &  \textbf{44} &  \textbf{5.9} & \textbf{1.76} & \textbf{130} & \textbf{30} \\
$\mathrm{U}=240$    & 0  & Y & Y & 25/12/2015 21:36:00 & 27/12/2015 08:36:00 &  35 &  9.8 & 3.04 & 175 & 45 \\
$\mathrm{U}=480$    & 0  & Y & Y & 25/12/2015 03:36:00 & 26/12/2015 08:36:00 &  29 & 16.1 & 3.98 &  198 & 49 \\
\hline
$\mathrm{U}=50$     & 250 & Y & Y & 27/12/2015 15:36:00 &        XX           &  XX &  XX  &  XX  &  XX & XX\\
$\mathrm{U}=100$    & 250 & Y & Y & 27/12/2015 06:36:00 & 31/12/2015 13:36:00 & 103 &  2.9 & 0.40 &  71 & 34 \\
$\textbf{U = 200}$  & \textbf{250} & \textbf{Y} & \textbf{Y} & \textbf{26/12/2015 09:36:00} & \textbf{27/12/2015 23:36:00} &  \textbf{38} &  \textbf{5.6} & \textbf{1.79} & \textbf{133} & \textbf{39} \\
$\mathrm{U}=400$    & 250 & Y & Y & 25/12/2015 12:36:00 & 26/12/2015 21:36:00 &  33 &  9.7 & 2.99 & 161 & 36 \\
$\mathrm{U}=800$    & 250 & Y & Y & 24/12/2015 18:36:00 & 25/12/2015 22:36:00 &  28 & 15.7 & 3.24 & 174 & 29 \\
\hline
$\mathrm{\Omega}=0.01$     & 0 & Y & Y & 26/12/2015 14:36:00 &      XX             & XX &  XX  &  XX &  XX &  XX \\
$\mathrm{\Omega}=0.03$     & 0 & Y & Y & 25/12/2015 19:36:00 & 27/12/2015 20:36:00 & 49 & 10.7 & 1.39 & 176 &  25 \\
$\textbf{$\Omega$ = 0.06}$ & \textbf{0} & \textbf{Y} & \textbf{Y} & \textbf{25/12/2015 07:36:00} & \textbf{26/12/2015 19:36:00} & \textbf{36} & \textbf{16.8} & \textbf{1.01} & \textbf{184}  & \textbf{19} \\
$\mathrm{\Omega}=0.12$     & 0 & Y & Y & 25/12/2015 01:36:00 & 26/12/2015 02:36:00 & 25 & 27.3 & 1.08 & 115 & 25 \\   
$\mathrm{\Omega}=0.24$     & 0 & Y & Y & 24/12/2015 17:36:00 & 25/12/2015 09:36:00 & 16 & 38.8 & 1.73 & 172 & 21 \\  
\hline
$\mathrm{\Omega}=0.03$     & 250 & Y & Y & 25/12/2015 22:36:00 & XX                  & XX &  XX  &   XX &  XX & XX \\
$\mathrm{\Omega}=0.07$     & 250 & Y & Y & 25/12/2015 02:36:00 & 26/12/2015 14:36:00 & 36 & 14.9 &   1.02 &  132 & 22 \\
$\textbf{$\Omega$ = 0.13}$ & \textbf{250} & \textbf{Y} & \textbf{Y} & \textbf{24/12/2015 14:36:00} & \textbf{25/12/2015 23:36:00} & \textbf{33} & \textbf{22.1} & \textbf{1.06} & \textbf{136} & \textbf{21} \\
$\mathrm{\Omega}=0.26$     & 250 & Y & Y & 24/12/2015 08:36:00 & 25/12/2015 06:36:00 & 23 & 32.6 & 2.01 & 203 & 30 \\   
$\mathrm{\Omega}=0.52$     & 250 & Y & Y & 24/12/2015 04:36:00 & 24/12/2015 19:36:00 & 15 & 41.6 & 2.02 & 263 & 23 \\  
\iffalse
                 \hline
E = 0 & 0    & N  & N        &     XX           &     XX         &    XX        & XX & XX & XX \\  
                 \hline
E = 0 & 250    & N  & N        &    XX            &     XX         &    XX        &  XX & XX & XX \\  
\fi
                 \hline
                 \hline
\end{tabular}

\end{sidewaystable*}

\section{Summary and Discussion}
\label{sec:section4}

We performed data-driven time-dependent magnetofrictional modelling (TMFM) of active region AR12473 and studied the effects of the values of $U$ and $\Omega$ used in determining the non-inductive photospheric electric fields on the formation and early evolution of the flux ropes and on some of their derived key parameters.  %Two free parameters are the $U$- and $\Omega$ parameters, 
%related to the 
%the vertical rise speed of the twisted flux tube
%to the photosphere and the horizontal component
%of fluid vorticity at the photosphere, respectively. 
First the optimised values of $U$- and $\Omega$ parameters were found by comparing photospheric energy injection with the energy injection from DAVE4VM. The simulations were performed with $[0.25,0.5,1,2,4] \times$ the optimised values and both for the masked (threshold $B = 250 G$) and unmasked magnetograms. 

The key findings can be summarized as follows:

\begin{itemize}
    \item Flux ropes formed in every simulation where non-inductive component was not zero. They also had overall similar appearance and evolution with the largest difference being in the formation and evolution times. For the runs made with setting the non-inductive component to zero the flux rope did not form at all (both for masked and unmasked magnetograms) during the simulation time. This is consistent with previous studies \citep{Cheung2012,Pomoell2019} and emphasizes the importance of including the non-inductive electric field in the TMFM simulations.
    \item Flux ropes formed earlier and evolved considerably faster in the $\Omega$ runs than in the $U$-runs using the same scaling value of the free parameters with respect to their optimized values. This could be related to significantly higher photospheric helicity injection (an order of magnitude) for $\Omega$ runs than for the $U$ presumably due to strong twisting motion it presents. In contrast, the photospheric energy injection for the corresponding electric fields were similar. 
    \item Flux ropes formed earlier and rose faster in the masked simulations than in unmasked masked simulations. The difference was however smaller than between the $\Omega$ and $U$ simulation runs. Photospheric helicity injection was found to be larger for masked cases than for unmasked cases, while no difference was found between the energy injections. 
    \item The flux rope formed increasingly earlier and rose increasingly faster with the increasing values of free parameters in the non-inductive electric field. It is not straightforward to match the flux rope evolution in the simulation with the real eruption time. For the studied event the flux ropes formed at the bottom of the simulation domain considerably earlier than the real eruption time (from 1-4 days). The height of $\approx 100 ~\textrm{Mm}$ was also reached earlier than the real eruption time for most of the runs. 
    %For optimised $U$ simulation for optimised $U$ unmasked 
    %All the flux ropes were evolving in size and height with time. 
    %\item We obtained the optimisation values by optimising the injection of magnetic energy. We then varied these values by $1/4 \times, 1/2 \times, 1 \times, 2 \times \& ~4 \times $ of the optimised value. 
    %\item We also studied the masking effect (threshold $B = 250 G$). We found that for the masked simulation, the flux ropes had appeared earlier than the unmasked runs. 
    \item The  current-carrying helicity increases, reaches maximum just before the eruption/when it reaches the height $\approx 100 ~\textrm{Mm}$ and then it decreases with time. 
    %\item In $\Omega$ simulations, the flux ropes appeared almost at the beginning and erupted much faster than the $U$ runs.
    %\item The higher the optimisation parameter values, the faster the flux ropes had formed and erupted. 
    \item The accumulated helicity, magnetic fields and axial flux were always higher for $\Omega$ simulation than $U$. It was also noted that increasing the optimisation values, increases all these parameters for $\Omega$ and $U$ runs.
    \item The size of the flux ropes (in terms of radius) were also increased with higher value of $\Omega$ and $U$ in most of the runs.
    \item The twist was found consistently positive for all flux ropes. The derived parameters (accumulated helicity, axial flux and current-carrying helicity) varied between the simulations but were of the same order of magnitude.
\end{itemize}

 The results of this study indicate that flux rope is formed and has overall similar evolution and properties with a large range of \textit{ad hoc} free parameters needed to determine the non-inductive electric field component that is  critical for energizing and introducing twist to the coronal magnetic field. As discussed in the Introduction the flux rope  parameters derived from the TMFM simulation can be used to constrain magnetized CME flux ropes models inserted in heliospheric simulations and to semi-empirical CME models. However, this study shows that, irrespective of the values, flux ropes are formed and erupted. Therefore data-driven TMFM can be used to estimate flux rope properties early in their evolution without needing to employ a lengthy optimisation process.

\begin{acknowledgements}
The authors acknowledge the European Research Council (ERC) under the European Union's Horizon 2020 Research and Innovation Programme Project SolMAG 724391 and the Academy of Finland Project 310445. This research used version 0.0.7 (software citation) of the SunPy open source software package.
\end{acknowledgements}

%\bibliographystyle{aa} % style aa.bst
%\bibliography{simulation_bib} % your references Yourfile.bib

\end{document}